\documentclass{elsart}

\def\be{\begin{equation}}
\def\ee{\end{equation}}
\def\bea{\begin{eqnarray}}
\def\eea{\end{eqnarray}}

\usepackage{graphicx,amssymb}

\begin{document}

\begin{frontmatter}

\title{Quantum Magnets with Anisotropic Infinite Range Random Interactions.}
\author{Liliana Arrachea$^{1,2}$ and Marcelo J. Rozenberg$^{1,3}$.}
\address{$^{1}$  Departamento de F\'{i}sica, FCEN, Universidad de Buenos Aires, Ciudad Universitaria Pab.1,
1428 Buenos Aires, Argentina.\\
$^{2}$ Instituto de Biocomputaci\'on y F\'{\i}sica de Sistemas Complejos (BIFI),
Corona de Arag\'on 42, Zaragoza 50009, Spain.\\
$^{3}$ Laboratoire de Physique des Solides,
Universit\'e de Paris-Sud, Orsay 91405, France.}

\begin{abstract}
Using exact diagonalization techniques we study the dynamical response
of the anisotropic disordered Heisenberg model for systems of $S=1/2$ spins
 with infinite range random exchange interactions at temperature $T=0$. 
The model can be considered as a generalization, to the quantum case, of the
well known Sherrington-Kirkpatrick classical spin-glass model.
We also compute and study the behavior of the 
Edwards Anderson order parameter and energy per spin 
as the anisotropy evolves from the Ising to the 
Heisenberg limits.  \end{abstract}


\medskip


\end{frontmatter}

Many real materials display a spin-glass phase at low enough temperature.
This exotic state is characterized by a frozen configuration of local magnetic 
moments following a random spatial pattern, in such a way that no net macroscopic
magnetization is produced. 

The manifestations of this state were first observed in the seventies in systems with
magnetic impurities in a metallic host (examples are Cu-Mn, Ag-Mn, Au-Mn and Ag-Mn).
There the magnetic moments experiment long range RKKY interactions, but spin-glass
behavior was also subsequently observed in the insulating compound Eu$_x$Sr$_{1-x}$S with 
competing ferromagnetic 
and random antiferromagnetic exchange interactions \cite{reviews}. 
More recently, spin glass phases have also been observed in the bi-layer kagome   
SrCr$_8$Ga$_4$O$_{19}$ \cite{uemu}, in the pyrochlore structure Li$_x$Zn$_{1-x}$V$_2$O$_4$
\cite{ura}, in the dipolar magnet LiHo$_x$Y$_{1-x}$F$_4$ \cite{liho} and in the enigmatic
high Tc compounds La$_{1-x}$Sr$_x$Cu$_2$O$_4$ \cite{elbio}. 

Competing interactions and frustration in combination with randomness have been 
identified as the basic ingredients to drive a system towards a glassy state.
While the concept of spin is purely quantum, it is usually stated that 
quantum fluctuations are not important to describe the spin glass physics. However,
the relevance of quantum effects is beginning to be identified and emphasized in 
experimental \cite{kado,qan,aeppli,isvshei} and theoretical work \cite{nos1,nos2,you,alberto}.

Exact diagonalization (ED) techniques proved to be very useful
and efficient to investigate the dynamical properties of strongly correlated systems
in general \cite{elbio}. In recent works we showed that it is also a useful method
to deal with quantum random spin systems with {\em long range interactions},
such as the Ising model with randon exchange interaction in the presence of a 
transverse and longitudinal magnetic fields \cite{nos1} and 
the random Heisenberg model \cite{nos2}. Most of the analytical and numerical methods to
study these kind of systems rely on the so called replica-trick. Unfortunately, this
clever technique becomes usually impractical within the glassy phase, where replica symmetry
breaking occurs. As ED does not use replicas, this technical difficulty is not encountered, and
the paramagnetic and glassy phases can be studied in the same way. Another appealing feature of ED 
is that it allows for the direct calculation of the dynamical response on the real frequency axis. 
In this way, it circumvents the uncertainties related to the 
 analytical continuation procedures which are usually
needed in quantum Monte Carlo simulations \cite{grempel}. Finally, 
another important advantage of ED
is the possibility of gaining insight on the nature of the low energy excitations.
The price to pay is that only small systems
 are amenable to be treated within the available computer power. 
However, at least for the case of models with 
infinite range interactions, in most of the cases the relevant physical quantities were 
found to extrapolate smoothly to the thermodynamic limit and a consistent description of
the different phases has been possible \cite{nos1,nos2}.   

In this work we present results on the behavior of the anisotropic Heisenberg model for a system
of $N$ spins with $S=1/2$ and random infinite-range exchange interactions. The
Hamiltonian reads,
\begin{equation}
H= \frac{1}{\sqrt{N}}\sum_{i,j=1}^N J_{ij} [S^z_i S^z_j +  \frac{\alpha}{2} (S^+_i S^-_j +
S^-_i S^+_j)], 
\label{hamil}
\end{equation}
where $i,j$ label the sites of the fully connected lattice and the interactions 
$J_{i,j}$ are normally distributed with variance $J^2$ that we set to unity. The parameter
$\alpha$ labels the degree of anisotropy interpolating between the classical Sherrington-Kirpatrick 
(SK)
model for $\alpha=0$ 
and the Heisenberg model for $\alpha=1$.
This model has a rather close experimental realization in the above mentioned
LiHo$_x$Y$_{1-x}$F$_4$ \cite{liho} compound in the limit of small magnetic ion
concentration $x$, which is dominated by random exchange interactions. The magnetic
interaction in this system is strongly anisotropic \cite{aeppli}.
We calculate the different components of the 
local dynamical susceptibility, defined as
\begin{equation}
\chi_{\mu}^{loc}(\omega) = \frac{1}{M}\sum_{M=1}^{m} 
\frac{1}{N}\sum_{i=1}^{N} 
\langle \Phi_0^{(m)}| S^{\mu}_i\frac{1}{\omega- H^{(m)}}S^{\mu}_i|\Phi_0^{(m)}\rangle,
\label{chis}
\end{equation} 
where $m$ denotes the number of realizations of disorder and
$\mu=x,y,z$ . The 
state $|\Phi_0^{(m)}\rangle$ is the ground state of the
Hamiltonian $H^{(m)}$, defined for the $m$th realization of $J_{ij}$ and it is calculated by
recourse to Lanczos algorithm. The evaluation of the spectral functions 
$\chi''_{\mu}(\omega)=-2\mbox{Im}[\chi_{\mu}^{loc}(\omega)]$ is achieved by 
using a continuous fraction representation of the dynamical response functions \cite{elbio}. 
We perform averages over a number of realizations of disorder between $M=3000$ for the largest systems 
($N=15$ spins) to $M=100000$ for the smallest ones ($N=8$ spins). 

\begin{figure}
\begin{center}
\includegraphics*[width=70mm,angle=-90]{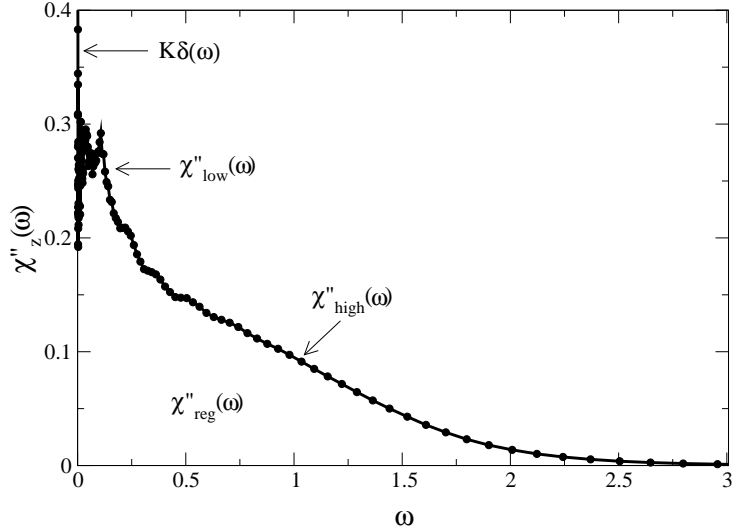}
\end{center}
\vspace{0.25cm}
\caption{The spectral function $\chi''_z(\omega)$ for the Heisenberg model ($\alpha=1$)
in a system of $N=14$ spins. The arrows indicate contributions from different kind of
excitations (see text).}
\label{Fig1}
\end{figure}

Let us first summarize the main properties of the spectral function in 
 the limit $\alpha=1$ (usual Heisenberg model) where the three components of the
 susceptibilities coincide \cite{nos1,nos2}. This is shown in Fig.1.
Four different contributions to the dynamical response can be distinguished:
\begin{equation}
 \chi''_z(\omega)=K\delta(\omega)+\chi''_{low}(\omega)+
\chi''_{high}(\omega)+ \chi''_{reg}(\omega)
\label{cont}
\end{equation}

The pure delta-function of the first term is a consequence of the
SU(2) rotational invariance of the Hamiltonian in this limit. This response
is due to a ``soft mode'' present in disorder realizations
with total $S\neq 0$ which have $2S+1$ fold degenerate ground states. A close analysis   
of the finite size effects on the average magnetization per site indicates that
 $<S>/N \rightarrow 0$ as the size of the system evolves to the thermodynamic limit. Hence,
this contribution to the spin response is expected to vanish as $N$ increases.

The analysis of the low frequency feature
$\chi''_{low}(\omega)$ as $N$ grows reveals that this contribution of the spectral function 
gets sharper, evolving towards a function $\sim \delta(\omega)$. Its origin 
has been associated to slow coherent excitations of many-spin states which
eventually become frozen. These states bear some resemblance to the magnons of the model 
without disorder. In fact, by examination of the structure of the
ground-state in typical realizations of disorder one finds that these excitations 
are built-up on a ground state whose wave 
function has large amplitudes on just a few configurations (out of the
2$^N$ states) corresponding to {\em unfrustrated sub-clusters}, i.e. a sub-set
of spins that are in a configuration that is compatible with the sign of the 
random $J_{ij}$ bonds.
Such configurations appear in
pairs together with their time-reversal counterparts, having the same weight in the ground state
 wave function and some relative sign that depends on the total $S$ of the ground state.  
The wave functions of excited states  that contribute to $\chi''_{low}(\omega)$  also
have a large weight on the unfrustrated clusters where the pairs of configurations
appear with the opposite relative sign with respect to the ground state. The typical
energy scale for these lowest energy excitations is $O(J/N)$.
The high-energy part $\chi''_{high}(\omega)$ is a small contribution with
the form of a mild hump which is produced by excitations generated from the 
unbinding of single spins out of the unfrustrated clusters,
the classical picture being a  
precession of individual spins around the effective quasi-static local
field of the remaining frozen ones (of the unfrustrated sub-cluster).
It is remarkable that the three above mentioned features, which are the consequence of
some kind of magnetic order are mounted on a broad and large background $\chi''_{reg}(\omega)$
that  remains almost unaffected as the size of the system increases.
This piece of the response
has been identified as due to {\em incoherent} excitations.  

\begin{figure}
\begin{center}
\includegraphics*[width=80mm,angle=-90]{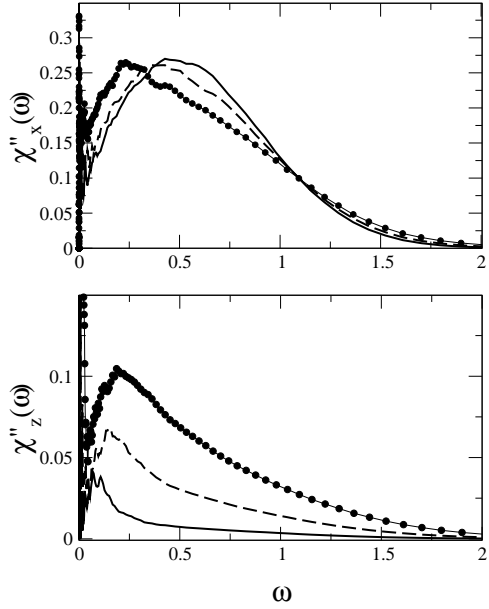}
\end{center}
\vspace{0.25cm}
\caption{The spectral function $\chi''_x(\omega)$ (upper panel) and $\chi''_z(\omega)$
(lower panel) for $\alpha=0.2,0.4,0.6$ (bold line, dashed line, circles).
in a system of $N=14$ spins.}
\label{Fig2}
\end{figure}

With this picture in mind, let us  now  carry out a similar analysis on the model
with anisotropy. For $\alpha \neq 0$, SU(2) rotational invariance
is broken and only the rotational invariance with respect to the $z$ axis 
is preserved, so that $\chi''_{x}(\omega) = \chi''_{y}(\omega)\neq
\chi''_{z}(\omega)$. The behavior of these spectral functions for different values of the
parameter $\alpha$ is shown in Fig.2. Both spectral functions contain a piece of the
form $K \delta(\omega)$. Its origin is the degeneracy of ground states with large
$S_z$ (or $S_x=S_y$ components). In the case of  $\chi''_{x}(\omega)$, this piece
carries
a very low spectral weight and, in both cases, it is observed to decrease as the size of
the system grows. Therefore, as in the pure Heisenberg case, this contribution is
expected to vanish in the thermodynamic limit. Save from this feature, 
$\chi''_{x}(\omega)$ does not show any clear indication of glassy order along this direction.
The remaining part of the spectral function behaves as produced by an incoherent continuum,
experimenting negligible changes when $N$ is modified, {\em without developing 
low energy features}. We identify it with the regular part 
$\chi''_{x,reg}(\omega)$. The spectral function corresponding to the response 
along the $z$ direction also contains some portion of its weight distributed on
a wide range of frequencies that is interpreted as caused by incoherent excitations.
Interestingly, at least for strong enough anisotropy,  $\chi''_{x,reg}(\omega)$ and
$\chi''_{z,reg}(\omega)$ seem to
behave linearly in $\omega$ at low frequency, 
as predicted by a mean-field description which is exact for
the SU(M) extension of the Heisenberg model in the limit of M$\rightarrow \infty$ 
\cite{alberto}. A more detailed study of the origin of this behavior
 is left for future investigations.

We now turn to analyze the contribution  $\chi''_{low}(\omega)$ which
is clearly distinguished within the low frequency region of $\chi''_{z}(\omega)$.
In an analogous way to what happens in the case of the isotropic Heisenberg model, the
typical width of this feature is $O(\alpha J /N)$ and evolves towards a $\sim \delta(\omega)$
as the system approaches the thermodynamic limit. Its origin is also due to excitations where
the spin configurations conforming unfrustrated clusters, which in the ground state
are frozen along the $z$ direction, become canted with a finite component
on the $xy$ plane in a coherent fashion. 
Assuming that the functional form of
the spectral function in the thermodynamic limit is $\chi''_{z}(\omega)=q\delta(\omega)
+\chi''_{z,reg}(\omega)$, the extrapolations  for the weight accumulated in 
$\chi''_{low}(\omega)$ as $N \rightarrow \infty$ provides an estimate of the
Edwards Anderson parameter $q$.

\begin{figure}
\begin{center}
\includegraphics*[width=80mm,angle=-90]{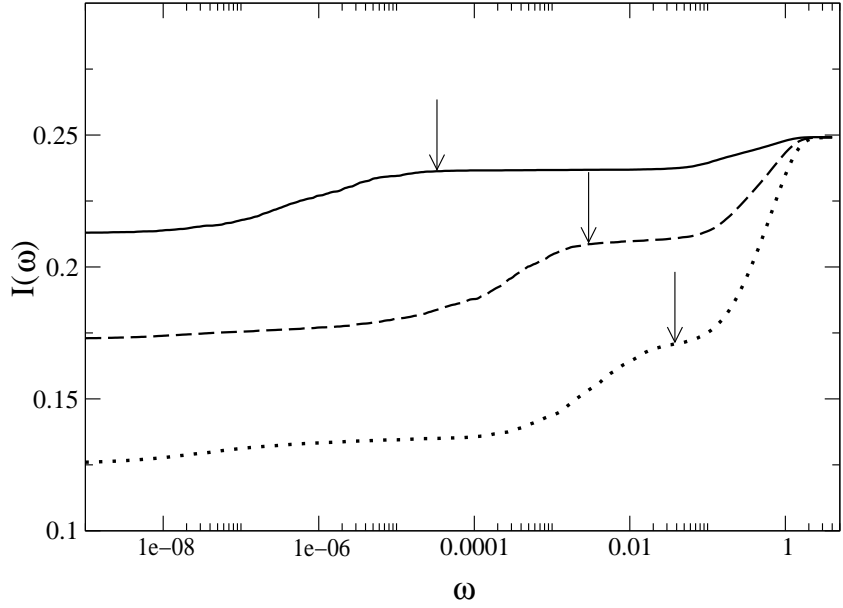}
\end{center}
\vspace{0.25cm}
\caption{The integrated spectral weight corresponding to $\chi''_z(\omega)$ 
for $\alpha=0.2,0.4,0.6$ (bold, dashed and dotted line) in a system of $N=14$ spins.
The arrows indicate the
weight accumulated in the low frequency piece $\chi_{low}(\omega)$.
}
\label{Fig3}
\end{figure}

Recalling that the sum rule for the spectral functions is
\begin{equation}
\int_{-\infty}^{+\infty} d\omega \chi''_{\mu}(\omega)=\frac{1}{4},
\label{sumrule}
\end{equation}
it is clear in Fig.2 that an important transfer of spectral weight takes place 
from $\chi''_{z,reg}(\omega)$ to $q$ as $\alpha$ approaches the classical limit
$\alpha=0$. A more quantitative picture of this effect is shown in the
logarithmic plot of the integrated weight $I(\omega)=\int_{-\infty}^{\omega} d\omega'
\chi''_{z}(\omega')$ shown in Fig.3. The weight accumulated in the low frequency peaks
defines plateaus in $I(\omega)$ that are indicated with arrows in the figure.
Extrapolations of the ordinates of these plateaus as the size of the system
increases provides us a reliable estimate of the Edwards Anderson parameter
$q$ in the thermodynamic limit. Results are shown in the upper panel of Fig.4.
This plot can be reasonably fitted with the function
$q=1/4\sqrt{1-(\alpha/\alpha_c)^2}$, with $\alpha_c \sim 1.02$. Interestingly enough,
this functional form suggests that, starting from the SK
model, the fluctuations introduced via the interaction between the $x,y$ components of 
the spins behave in a similar way to those driven by the temperature $T$ in the pure classical model.
In the latter case it has been found 
 $q-1/4 \propto T^2$ \cite{reviews}, for $T \sim 0$. Instead, it seems that the effect of 
quantum fluctuations  in the anisotropic Heisenberg model is different 
from those caused by adding a transverse field $\Gamma$ to the classical SK model, where
$q=1/4(1-\Gamma/\Gamma_c)$ \cite{physb}.  This kind of behavior should be relevant
in the analysis of quantum vs classical annealing recently studied experimentaly in
the LiHo$_x$Y$_{1-x}$F$_4$ compound \cite{qan,aeppli}.
The energy per site as a function of the inverse of the system size
is shown in the lower left panel of Fig.4. The result of the extrapolations
to the limit $N \rightarrow \infty$ are shown in the lower right panel.
For comparison, we also show the estimates provided by the replica symmetric solution \cite{sherkir}
and the one and two step replica symmetry breaking solutions \cite{parisi} for the classical model,
corresponding to $e=-0.1995, -0.1913,-0.1909$, respectively \cite{note}. 
It is worth to stress that, within that approximation,
 improving the level of symmetry breaking leads to higher values of the
ground state energy per site. 
We  have not explicitly investigated the limit $\alpha=0$. However, it is clear from Fig.4  that
 a simple monotonic continuation of the curve
depicted by our numerical results leads at this limit to $e \sim -0.183$, i.e. to a value even larger than that
predicted by the two-step replica symmetry breaking solution. As it is likely that higher levels 
of symmetry breaking give higher energies, we can say that our estimate provides 
an upper bound to the exact ground-state energy of the classical model.  

\begin{figure}
\begin{center}
\includegraphics*[width=80mm,angle=-90]{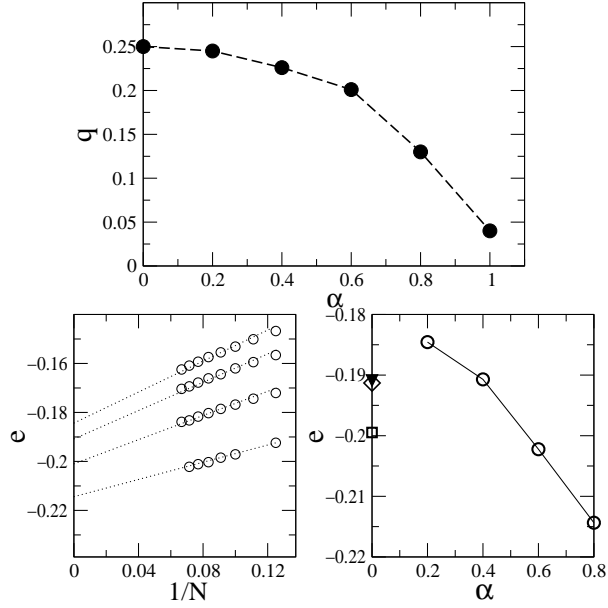}
\end{center}
\vspace{0.25cm}
\caption{Upper panel: The extrapolated values for the Edwards Anderson order parameter as a function of the
strength of anisotropy. Lower left panel: Energy per site as a function of the inverse of the
lattice size. Lower right panel: Extrapolated energy per site as a function of $\alpha$.  
The square, diamond and triangle correspond to the replica symmetric \cite{sherkir}, one and two step
replica symmetry breaking \cite{parisi} solutions.}
\label{Fig4}
\end{figure}

To conclude, we have investigated a model of a quantum spin-glass with long range anisotropic
interactions which is a quantum generalization of the Sherrington-Kirkpatrick model.
Our results show that both $\chi_z''(\omega)$ and $\chi_x''(\omega)$ have a rich 
structure. We found that the $z$-component remains frozen
for all values of the anisotropy and that the order parameter $q$ decreases very fast
for anisotropies $\alpha>0.6$, being remarkably weak in the pure Heisenberg limit.

It is worth mentioning that for, large enough anisotropy, the low frequency edge of the 
regular part of $\chi''_{x}(\omega)$ and $\chi''_{z}(\omega)$ 
seem to be consistent with linear behavior. This point
is in contrast with our previous analysis for the SU(2) isotropic model \cite{nos1,nos2}, 
but it is
in agreement with the results obtained in the large M limit
\cite{alberto}.

We acknowledge support from CONICET (Argentina) and ANPCyT PICT 03-11609 (Argentina). LA Thanks the support of the MCyT of Spain through the ``Ram\'on y Cajal''program.


\begin{thebibliography}{00}
\bibitem{note}There is a factor $1/4$ between our results for $e$ and $q$ in comparison to those
reported in the literature for the classical model. This is due to the fact that it is frequently
assumed $S^z_i=\pm 1$ for the classical model, while in our case $S^z_i=1/2$.
\end{thebibliography}
\end{document}